\documentclass[aps,prb,twocolumn]{revtex4}
\usepackage{amsmath}

\begin{document}
\title{Comment on "Poynting vector, heating rate, and stored energy in
  structured materials: A first principles derivation"}
\author{Vadim A. Markel}
\affiliation{Departments of Radiology, Bioengineering, and the
  Graduate Group in Applied Mathematics and Computational Science, University
of Pennsylvania, Philadelphia, Pennsylvania 19104} 

\begin{abstract}
  In this Comment I argue that Silveirinha's criticism of my earlier
  work [M.G.Silveirinha, Phys. Rev. B, {\bf 80}, 235120 (2009)] is
  based on an unphysical excitation model which involves an ``external
  current'' which overlaps with a continuous medium but is not subject
  to constitutive relations. When this excitation model is replaced by
  the conventional and experimentally-relevant model of excitation by
  external electromagnetic fields, it can be easily shown that
  Silveirinha's formulas contain the very results he wanted to
  disprove. This and a few other misconceptions present in
  Silveirinha's paper are subject of this Comment.
\end{abstract}

\date{\today}

\maketitle
\section{Introduction}

This paper contains a comment on the paper~\cite{silveirinha_09_1} by
M.G.~Silveirinha entitled ``Poynting vector, heating rate, and stored
energy in structured materials: A first-principles derivation'' [{\em
  Phys.~Rev.~B} {\bf 80}, 235120 (2009)].

As is customary for all APS journals, the Comment was first sent to
Silveirinha for review. Silveirinha has written an 11-page report on
my Comment and recommended its rejection.  I disagreed, the Comment
went through the ordinary process of peer-review and was eventually
rejected. Still, I did not agree with the scientific arguments used to
justify the rejection. I have then reworked the Comment into a
stand-alone paper with independent conclusions. This
paper~\cite{markel_10_2} is now published and is entitled ``On the
current-driven model in the classical electrodynamics of continuous
media'' [{\em J.  Phys.-Cond.~Matt.} {\bf 22}, 485401 (2010)].
Although the above publication makes all the essential points, I still
feel that the original Comment should be made available to the
community as it is, in some respects, more direct and specific.

Finally, it should be mentioned that a different
comment~\cite{richter_10_1} on the same paper~\cite{silveirinha_09_1}
by Silveirinha has been submitted and published.

\section{The original Comment}

Recently, Silveirinha has criticized~\cite{silveirinha_09_1} several
earlier papers~\cite{markel_08_1,richter_08_1,markel_08_2}, two of
which are authored by myself. In particular, Silveirinha has stated
that the conclusions made by me in Ref.~\onlinecite{markel_08_1}, as
well as similar conclusions made by Richter {\em et al.} in
Ref.~\onlinecite{richter_08_1}, "are founded on fundamental
misconceptions and mistakes". To support this claim, Silveirinha has
offered a rather lengthy derivation which, according to Silveirinha,
disproves the results of Refs.~\onlinecite{markel_08_1} and
\onlinecite{richter_08_1}. However, I have found that Silveirinha's
derivation itself is based on a few misconceptions. The chief among
these is the contention that an electromagnetic wave with a purely
real wave vector can propagate in absorbing materials (either "meta-"
or otherwise), that such non-decaying waves can be obtained by
imposing an ``external current'' which overlaps with the medium but is
not subject to constitutive relations, and that solutions thus
obtained can be used to derive information about the rate at which the
medium is heated by radiation (the heating rate). This and a few other
"misconceptions and mistakes" which are present in Silveirinha's work
are subject of this Comment.  Gaussian system of units is used
throughout. Equations of Silveirinha's paper are referenced as (S\#).

{\bf A) The assumption of purely real wave vector and the external
  current.}  Most of Silveirinha's derivations, including all
derivations that lead to results of any practical significance, are
carried out for Bloch waves with a purely real wave vectors ${\bf k}$.
Silveirinha is, of course, aware that natural Bloch modes in periodic
media with some amount of absorption are, necessarily, decaying waves.
In a typical experiment, these decaying modes are excited by external
radiation. The decay is mathematically represented by the imaginary
part of ${\bf k}$. Thus, if ${\bf k}$ is taken to be purely real, the
medium is, by definition, non-absorbing. Calculation of the heating
rate in such a medium is meaningless: under the condition ${\rm
  Im}{\bf k}=0$, any reasonable calculation must yield zero.

Silveirinha, however, contends in his paper that a real-valued Bloch
wave vector ${\bf k}$ is not incompatible with losses. He argues that
one can consider an {\em external current} rather than an external
radiation as the source for the electromagnetic wave in the medium.
This is quite fine as long as the current and the medium do not
overlap. But Silveirinha requires that the external current be of the
form ${\bf j}_e = {\bf A}\exp(i{\bf k}\cdot {\bf r})$, where ${\bf A}$
is a constant amplitude, so that ${\bf j}_e$ is nonzero in the region
of space occupied by the medium. It must be emphasized that the
current ${\bf j}_e$ introduced by Silveirinha is not created by the
charges associated with the medium, does not obey any constitutive
relations (and, therefore, is independent of the fields existing in
the medium), and, presumably, can be manipulated by the experimenter
at will.

The excitation model proposed by Silveirinha and described briefly
above is unphysical because it can not be realized experimentally. The
obvious question Silveirinha should have answered is the following:
how can an external current of the form ${\bf j}_e = {\bf A}\exp(i{\bf
  k}\cdot {\bf r})$ be physically created in the medium? If we set
aside heating, mechanical stress, acoustic waves and other similar
influences (which, anyway, are highly unlikely to generate the
required current), we are left with the possibility of applying
external electromagnetic fields to the medium. But any current
produced in such an experiment is subject to the usual constitutive
relations and is, therefore, a part of the induced current which is
denoted in Ref.~\onlinecite{silveirinha_09_1} by ${\bf j}_d$. Another
option is to insert multiple wires into the medium and to run
externally-controlled currents through the wires.  This will, of
course, destroy the medium and no reasonable experimentalist would
propose or try to implement such an excitation scheme. It is also
highly unlikely that pre-determined spatially-varying currents can be
maintained in the wires by external voltages.

Thus, the idea of exciting the medium with a pre-determined external
current which overlaps with the medium is unphysical and can not be
used to derive any useful physical quantity. But even from a purely
formal point of view, any results derived from such an excitation
model can not be used to criticize Ref.~\onlinecite{markel_08_1}
because in this paper I have clearly and repeatedly stated that I
consider only the case when all currents in the medium are induced and
there are no external or ``free'' currents within the medium. (Note
that the conductivity current is included in the induced current in
this paper and in Ref.~\onlinecite{markel_08_1}.)

{\bf B) Heating rate with and without the assumption of a purely real
  wave vector.}  Now let us consider how Silveirinha computes the
heating rate. The starting point will be Eq.~(S60). To save a few
notations, I will re-write this equation in Gaussian units and omit
all subscripts. Then Eq.~(S60) takes the form

\begin{equation}
\label{q1}
q= \frac{\omega}{8\pi}{\rm Im}\left[ {\bf E}^* \cdot
  \hat{\varepsilon} {\bf E} + \left(\frac{c}{\omega}\right)^2 {\bf E}^* \cdot {\bf
    k}\times (\hat{\mu}^{-1}-\hat{I}){\bf k} \times {\bf E} \right] \
.  
\end{equation}

\noindent
Here tensors (dyadics) are denoted by a hat, the symbol "$\cdot$" is
reserved for a dot product of two vectors, no special symbol is used
to denote the action of a tensor on a vector (for example,
$\hat{\varepsilon}{\bf E}$), and all operations are evaluated from
right to left; ${\bf E}$ is the macroscopic electric field and
$\hat{I}$ is the unit tensor. All fields are assumed to be
monochromatic and the common exponential factor $\exp(-i\omega t)$ is
suppressed.  Note that I have used the first identity in (S36) to
transform (S60) to the form (\ref{q1}). In fact, (\ref{q1}) is
equivalent to (S60) but is written using more conventional tensor
notations.

Silveirinha evaluates expression (\ref{q1}) as follows. He uses the
vector identity ${\bf a} \cdot({\bf b} \times {\bf c}) = ({\bf a}
\times {\bf b}) \cdot {\bf c}$ (equivalent to the second identity in
(S36)) to re-write the second term in the square brackets as

\begin{equation}
\label{t2}
\left( \frac{c}{\omega} \right)^2
({\bf E}^* \times {\bf k}) \cdot [(\hat{\mu}^{-1}-\hat{I}){\bf k}
\times {\bf E}] \ .
\end{equation}

\noindent
Then Silveirinha uses Eq.~(S10a), namely, ${\bf k} \times {\bf E} =
(\omega/c) {\bf B}$. This equality follows from the macroscopic
Maxwell's equations applied to a plane wave. It can be seen that the
factor ${\bf k} \times {\bf E}$ in the end of expression (\ref{t2})
can be replaced by $(\omega/c) {\bf B}$. However, the expression ${\bf
  E}^* \times {\bf k}$ in the beginning of this expression can only be
replaced by $-(\omega/c) {\bf B}^*$ if ${\bf k}$ is purely real.
Silveirinha makes this assumption about ${\bf k}$ and transforms
(\ref{t2}) to the form

\begin{equation}
\label{t22}
-{\bf B}^* \cdot (\hat{\mu}^{-1}-\hat{I}){\bf B}
\end{equation}

\noindent
The term proportional to $\hat{I}$ is then omitted since its imaginary
part evaluates to zero, the field ${\bf B}$ is expressed in terms of
the field ${\bf H}$ using the constitutive relation ${\bf B} =
\hat{\mu}{\bf H}$ and Silveirinha arrives at his Eqs.~(S61) and (S62).
The important point here is that the transition from (S60) to (S61)
and (S62) requires that ${\rm Im}{\bf k} = 0$.

But if we set aside the unphysical model in which the medium is
excited by an external current which overlaps with the medium, then
${\bf k}$ is not a mathematically independent variable but is
determined from the Maxwell's equations and constitutive relations.
The equality ${\rm Im}{\bf k}=0$ is only possible if ${\rm
  Im}\hat{\varepsilon} = {\rm Im}\hat{\mu} = 0$. Therefore, the final
result for the heating rate derived by Silveirinha, Eq.~(S62),
evaluates to zero under the assumption that was used to derive it. It
can be concluded that the result (S62) is not a "first-principle,
completely general" derivation of some useful physical quantity, as
Silveirinha has claimed, but a lengthy and laborious proof of the
identity $0=0$.

It is possible, however, to evaluate the expression (\ref{q1})
differently without making any assumptions about ${\bf k}$. To this
end, we use the two Maxwell's equations

\begin{equation}
\label{Maxwell}
{\bf k} \times {\bf E} = \frac{\omega}{c} \hat{\mu}{\bf H} \ , \ \ {\bf k} \times {\bf H} = -\frac{\omega}{c} \hat{\varepsilon}{\bf E} \ .
\end{equation}

\noindent
Here I have assumed that all currents in the medium are induced and
subject to constitutive relations. From (\ref{Maxwell}), we can also
obtain

\begin{equation}
{\bf k} \times \hat{\mu}^{-1} {\bf k}\times {\bf E} =
-\left(\frac{\omega}{c} \right)^2 \hat{\varepsilon} {\bf E} \ .
\end{equation} 

\noindent
Substitute this expression into (\ref{q1}). The term proportional to
$\hat{\varepsilon}$ will cancel to yield

\begin{equation}
\label{q2}
q = -\frac{\omega}{8 \pi} \left(\frac{c}{\omega}\right)^2{\rm Im}
\left( {\bf E}^* \cdot {\bf k} \times {\bf k} \times {\bf E} \right) \
. 
\end{equation}

\noindent
Next, use the identity ${\bf k} \times {\bf k} \times {\bf E} = {\bf
  k} ({\bf k}\cdot {\bf E}) - k^2 {\bf E}$ to obtain

\begin{equation}
\label{q3}
q = \frac{\omega}{8\pi} \left(\frac{c}{\omega}\right)^2 {\rm Im}
\left[ \vert {\bf E} \vert^2 k^2 - ({\bf E} \cdot {\bf k}) ({\bf E}^*
  \cdot {\bf k}) \right] \ . 
\end{equation}

\noindent
This expression is equivalent to the one derived by me earlier in
Ref.~\onlinecite{markel_08_1} (Eq.~(54) of this reference) for the
case of general anisotropic nonlocal media. If the medium is
isotropic, it can support only transverse waves whose wave number is
$k^2 = (\omega/c)^2 \varepsilon\mu$ with $\varepsilon$ and $\mu$ being
scalars. Then (\ref{q3}) is simplified to

\begin{equation}
\label{q4}
q = \frac{\omega \vert {\bf E} \vert^2}{8\pi}{\rm Im} (\varepsilon \mu) \ .
\end{equation}

\noindent
Again, this result was derived by me in Ref.~\onlinecite{markel_08_1}
(see Eq.~(32)).

Thus, it can be seen that Silveirinha's Eq.~(S60) contains, in fact,
the very results he wanted to disprove. The only reason Silveirinha
has obtained a formula which appears to be different from (\ref{q3})
or (\ref{q4}) is because he has used a method to evaluate his equation
(S60) which is only valid when ${\rm Im}{\bf k}=0$. My method of
evaluating (S60) makes no assumptions about ${\bf k}$. It is
applicable, in particular, when ${\rm Im}{\bf k}=0$. In this case,
(\ref{q3}) and (\ref{q4}), as well as Silveirinha's results (S61),
(S62), all evaluate to zero and are, in this sense, equivalent. But,
unlike Silveirinha's results, formulas (\ref{q3}) and (\ref{q4}) can
also be used in the physically interesting case of a complex wave
vector ${\bf k}$.

Silveirinha would, perhaps, argue that, for complex wave vectors,
(S60) is itself invalid because its derivation requires that the
second term in the right-hand side of (S58) be neglected. This term
was shown to be exactly zero when ${\rm Im}{\bf k}=0$. However, in
electromagnetically homogeneous media, this term is negligibly small
anyway. This is evident from the following consideration. The tensor
$\hat{G}_{p0}$ defined in (S57) is real and Hermitian as long as ${\bf
  k} \cdot ({\bf r} - {\bf r}^\prime) \ll 1$. Since the double
integration in (S58) is over the volume of the same elementary cell,
the maximum phase shift ${\bf k}\cdot ({\bf r} - {\bf r}^\prime)$ is
of the order of or smaller than $\vert{\bf k}\vert h$, where $h$ is
the characteristic cell size. For the medium to be considered
electromagnetically homogeneous, this phase shift must be negligibly
small. Therefore, the result of double integration in (S58) is purely
real with the same precision as the precision of the underlying
approximation. In other words, the errors that a made when the
original periodic medium is replaced by a spatially-uniform medium
with effective parameters are of the same order of magnitude as the
errors which are made when the last term in (S58) is neglected.

{\bf C) The lack of equivalence between magnetization and
  nonlocality.} Another serious misconception perpetrated by
Silveirinha is that, at high frequencies, magnetic response of matter
can be understood and is physically indistinguishable from nonlocality
in the dielectric response. To be fair, we should note that
Silveirinha did not come up with this misconception on his own.  The
misconception can be traced to the highly respected book by Landau and
Lifshitz~\cite{landau_ess_84} and was recently popularized by
Agranovich~\cite{agranovich_06_2}. However, it is elementary to see
that no such equivalence exists.

Indeed, consider the induced current in a medium with electric and
magnetic polarization. Let, for simplicity, the medium be isotropic
with scalar coefficients $\varepsilon$ and $\mu$. For monochromatic
fields, the induced current is

\begin{equation}
\label{j1}
{\bf J} = -i\omega {\bf P} + c \nabla \times {\bf M} \ ,
\end{equation}

\noindent
where ${\bf P} = \chi_e {\bf E}$ and ${\bf M} = \chi_m {\bf B}$,
$\chi_e = (\varepsilon-1)/4\pi$ and $\chi_m = (\mu - 1)/4\pi\mu$.
Traditionally (e.g., in Landau and Lifshitz's book or in Agranovich's
review), the expression (\ref{j1}) is evaluated for a plane wave with
the wave vector ${\bf k}$ by writing

\begin{equation}
\nabla \times {\bf M} = \nabla \times \chi_m {\bf B} = \chi_m \nabla \times {\bf B} =
i\frac{c}{\omega} \chi_m {\bf k} \times {\bf k} \times {\bf E} \ .
\end{equation}

\noindent
Thus, it is found that the current can be equivalently written as

\begin{equation}
\label{J_nonl}
{\bf J} = -i\omega \frac{\hat{\epsilon}_{\rm nonl}(\omega,{\bf k}) -
  \hat{I}}{4\pi} {\bf E} \ ,
\end{equation}

\noindent
where

\begin{equation}
\label{eps_nonl}
\hat{\epsilon}_{\rm nonl}(\omega,{\bf k}) = \varepsilon \hat{I} -
\left(\frac{c}{\omega}\right)^2 \frac{\mu-1}{\mu} {\bf k} \times {\bf
  k} \times \ \ .
\end{equation}

\noindent
Upon observing that exactly the same induced current is obtained in a
medium with permeability equal to unity and nonlocal permittivity
$\hat{\epsilon}_{\rm nonl}(\omega,{\bf k})$ as the current obtained in
a medium with local permittivity $\varepsilon \neq 1$ and local
permeability $\mu \neq 1$, the conclusion is drawn about the physical
indistinguishability of the effects of magnetization and nonlocality
of the dielectric response.

The above line of reasoning is erroneous because $\nabla \times \chi_m
{\bf B} \neq \chi_m \nabla \times {\bf B}$. The correct formula is

\begin{equation}
\nabla \times \chi_m {\bf B} = \chi_m \nabla \times {\bf B} + (\nabla
\chi_m) \times {\bf B} \ .
\end{equation}

\noindent
The last term in the above equation is often lost sight of because it
is assumed that $\chi_m = {\rm const}$ and, consequently, $\nabla
\chi_m = 0$. However, even if the medium is quite uniform spatially,
it always has a boundary, and at the boundary the function
$\chi_m({\bf r})$ experiences a jump. Differentiation of this jump
gives rise to a surface current which can not be accounted for in the
model of nonlocal permittivity (\ref{eps_nonl}). To be more precise,
accounting for the surface current within the spatial nonlocality
model would require the use of position-dependent and highly singular
integral kernels in the influence function which would contain
derivatives of the delta function. This possibility is never
considered. In particular, in
Refs.~\onlinecite{landau_ess_84,agranovich_06_2}, the surface current
is simply ignored.

Thus, the deficiency of the equivalence model is that it
places the emphasis on the dispersion relation but ignores the medium
boundary. Although infinite boundless media can be considered as
purely mathematical abstractions, such media do not exists in nature.
And a dispersion relation alone does not characterize a finite-size
sample: the latter also has an impedance. Indeed, the dispersion
relation which follows from (\ref{J_nonl}),(\ref{eps_nonl}) is the
familiar equation $k^2 = (\omega/c)^2\varepsilon \mu$, where I have
used the fact that waves in isotropic media are transverse. This
dispersion relation is invariant with respect to the transformation
$\varepsilon \rightarrow \beta \varepsilon$, $\mu \rightarrow
\beta^{-1} \mu$, where $\beta$ is an arbitrary complex constant. In
particular, we can take $\beta = \mu$. It then would appear that a
medium with some values of $\varepsilon$ and $\mu$ is
indistinguishable from a medium with $\varepsilon^\prime =
\varepsilon\mu$ and $\mu^\prime = 1$. But the two media have different
impedances and, therefore, different transmission and reflection
properties.

{\bf D) The meaning of homogenization.} Yet another misconception
encountered in Silveirinha's work is related to the notion of
homogenization. Silveirinha seems to believe that homogenization is
equivalent to field averaging. However, fields can always be averaged
but not every medium is electromagnetically homogeneous at a given
frequency. The goals of homogenization are the following. Firstly, one
needs to prove that solutions to the macroscopic Maxwell's equations
in a nonmagnetic spatially-inhomogeneous medium characterized by the
``true'' permittivity $\varepsilon_{\rm true}({\bf r})$ converge in
some physically reasonable norm to solutions in a medium of the same
overall shape but with spatially-uniform effective parameters
$\varepsilon_{\rm eff}$ and $\mu_{\rm eff}$. Secondly, the effective
parameters must be computed explicitly in terms of $\varepsilon_{\rm
  true}({\bf r})$. The original and the effective media must be
``electromagnetically similar'' for varying incident fields and
varying shapes of the sample.

None of this has been done in Ref.~\onlinecite{silveirinha_09_1}.
Silveirinha did not consider the conditions under which spatial
averaging of fields is physically meaningful, nor did he derive the
effective parameters in terms of $\varepsilon_{\rm true}({\bf r})$.
True, more was done by Silveirinha in
Ref.~\onlinecite{silveirinha_07_1}, where a numerical procedure to
determine the effective parameters was proposed and implemented.
However, this development was not used in
Ref.~\onlinecite{silveirinha_09_1}. Besides,
Ref.~\onlinecite{silveirinha_07_1} is affected by the same unphysical
excitation model as Ref.~\onlinecite{silveirinha_09_1}: the Bloch wave
number ${\bf k}$ is viewed in both references as a
mathematically-independent variable ``imposed'' by the external
current although it is well known that to solve a photonic crystal,
${\bf k}$ must be computed self-consistently. In any case,
Silveirinha's derivation presented in
Ref.~\onlinecite{silveirinha_09_1} is simply a chain of definitions
which, after some terms are neglected, result in familiar macroscopic
formulas, such as (S60) (which Silveirinha then evaluates
incorrectly). These formulas contain effective parameters which, at
that point, are purely phenomenological and which have not been
computed either ``from first principles'' or in any other meaningful
way. For example, in Eq.~(S60), there appear some effective medium
parameters $\varepsilon_r$ and $\mu_r$. How are these related to
$\varepsilon_{\rm true}({\bf r})$? How does Silveirinha know that
$\mu_r$ is different from unity?

In this respect, an interesting result has been recently reported by
Menzel {\em et al.}~\cite{menzel_10_1} It was found that a typical
``metamaterial'' can not be reasonably characterized by effective
parameters in the spectral region in which magnetic resonances are
excited. This result indicates that, contrary to Silveirinha's
assertion, ``artificial magnetism'' can not be obtained in
electromagnetically homogeneous materials. Here, following
Ref.~\onlinecite{pendry_99_1} and a naive analogy with natural
magnetism, Silveirinha has confused magnetic polarizability of
subwavelength particles with magnetic permeability of a periodic
medium made of such particles. In the case of natural magnetism, it is
true that magnetically polarizable atoms can assemble to form a medium
with nontrivial permeability. However, natural magnetism is a quantum
effect which does not disappear at zero frequency. The so-called
artificial magnetism is a conceptually different and a purely
classical phenomenon. In particular, it should be noted that while the
magnetic polarizability of a particle is always defined, and can,
indeed, be large if Ohmic losses are sufficiently small, the effective
permeability of a medium is not always defined. The results of
Ref.~\onlinecite{menzel_10_1} indicate that the effective parameters
can only be introduced far away from magnetic resonances.

Silveirinha's correctly writes that the quadratic fluctuations of the
field of the type $\langle \delta{\bf e} \times \delta {\bf b}
\rangle$ ({\bf e} and {\bf b} are the ``microscopic'' electric and
magnetic fields) can not be neglected in the {\em general case}.
However, in Ref.~\onlinecite{markel_08_1}, I was not concerned with
the general case but only with the case of electromagnetically
homogeneous media which can be reasonably characterized by effective
medium parameters. In such media, the quadratic fluctuations can be
and should be, in fact, neglected. Accounting for such fluctuations
amounts to exceeding the precision of the underlying approximation
(that is, the approximation in which the medium is characterized by
spatially-uniform effective parameters). Silveirinha did not
demonstrate that the quadratic fluctuations can be large while the
medium is still electromagnetically homogeneous according to the
criteria stated above. And the quantitative arguments given in this
Comment are quite indicative of the contrary.

{\bf E) Unsubstantiated criticism.} Silveirinha's criticism of
Ref.~\onlinecite{markel_08_2} which appears after Eq.~(S62) is not
based on any scientific arguments. To be more specific, Silveirinha
has derived (S62) and stated, essentially, that it follows from (S62)
that Ref.~\onlinecite{markel_08_2} is erroneous as shown by Efros.
However, the argument of Ref.~\onlinecite{markel_08_2} is based on the
traditional expression for the heating rate (the one equivalent to
(S62)) and the alternative expression is only mentioned as a
possibility. Also, the arguments of the {\em earlier} paper by
Efros~\cite{efros_04_1} are considered and discussed in
Ref.~\onlinecite{markel_08_2} at length. Finally, the main subject of
Ref.~\onlinecite{markel_08_2} are natural diamagnetics. It almost
appears that Silveirinha did not look at the content of
Ref.~\onlinecite{markel_08_2} at all.

{\bf In summary}, I have shown that the main conclusions of
Silveirinha's paper are based on an unphysical excitation model in
which an external current which overlaps with the medium but is not
subject to constitutive relations is used. However, if we use instead
the conventional and experimentally-relevant model of excitation by
external fields, we would find that Silveirinha's formulas contain the
very results he wanted to disprove. I have demonstrated this
explicitly for the heating rate, $q$. I did not consider the density
of electromagnetic energy as this quantity is not measurable
experimentally and is, therefore, irrelevant (only the total
electromagnetic energy of a body is measurable, but this quantity is
the same in the conventional theory and in the theory of
Ref.~\onlinecite{markel_08_1}). As for the Poynting vector, ${\bf S}$,
in a steady state and, particularly, for monochromatic fields, it must
satisfy the continuity equation $q + \nabla \cdot {\bf S} = 0$ (time
averaging is assumed). I have shown that the quantity $q$ that follows
from Silveirinha's Eq.~(S60) is exactly the same as I have derived in
Ref.~\onlinecite{markel_08_1}. I have also shown in
Ref.~\onlinecite{markel_08_1} that the expression for ${\bf S}$ which
contains the cross product ${\bf E} \times {\bf H}$ is not consistent
with this expression for $q$ while the expression which contains ${\bf
  E} \times {\bf B}$ is. Therefore, if Silveirinha adopts the correct
excitation model and uses it to re-compute the Poynting vector, he
would surely find an expression which is quite in agreement with my
previous results and with Silveirinha's own Eq.~(S60).

\bibliographystyle{prsty}
\bibliography{abbrev,master,book,local}

\end{document}